\documentclass[10pt,conference]{IEEEtran}

\ifCLASSOPTIONcompsoc
  \usepackage[nocompress]{cite}
\else
  \usepackage{cite}
\fi

\usepackage{graphicx}
\usepackage{amsmath}
\usepackage[table,xcdraw]{xcolor}
\usepackage{booktabs}
\usepackage{siunitx}
\usepackage[linesnumbered,ruled,vlined]{algorithm2e}
\ifCLASSINFOpdf
\else
\fi

\begin{document}
%
\title{Security Assessment of  Interposer-based Chiplet Integration}

\author{\IEEEauthorblockN{Mohammed Shayan, Kanad Basu, Ramesh Karri}
\IEEEauthorblockA{New York University, \\
Brooklyn, NY 11201 \\ 
\{mos283,kb150,rkarri\}@nyu.edu}}
\maketitle

\begin{abstract}
With transistor scaling reaching its limits, interposer-based integration of dies (chiplets) is gaining traction \cite{yazdani2015inter}. Such an interposer-based integration enables finer and tighter interconnect pitch than traditional system-on-packages \cite{ssiyer20153d} and offers two key benefits: 1. It reduces design-to-market time by bypassing the time-consuming process of verification and fabrication. 2. It reduces the design cost by reusing chiplets. While black-boxing of the slow design stages cuts down the design time, it raises significant security concerns. We study the security implications of the emerging interposer-based integration methodology. The black-boxed design stages deploy security measures against hardware Trojans, reverse engineering, and intellectual property piracy in traditional systems-on-chip (SoC) designs and hence are not suitable for interposer-based integration. We propose using functionally diverse chiplets to detect and thwart hardware Trojans and use the inherent logic redundancy to shore up anti-piracy measures. Our proposals do not rely on access to the black-box design stages. We evaluate the security, time and cost benefits of our plan by implementing a MIPS processor, a DCT core, and an AES core using various IPs from the Xilinx CORE GENERATOR IP catalog, on an interposer-based Xilinx FPGA.
\end{abstract}


\IEEEpeerreviewmaketitle

\section{Introduction}
Packages have scaled very little when compared to transistors over the last few decades~\cite{ssiyer2016hetero}. In fact, while transistors have scaled 1000x, packages have scaled \~ 4x. Recently there is a trend towards further scaling of the  packaged product for three main reasons~\cite{yazdani2015inter}: 1) the transistor scaling reaching its limits; new technology nodes are not yielding the same power and area savings as at earlier nodes. 2) the demand for the integration of heterogeneous components fabricated on different technology nodes, and 3) the increase in Non-Recurring Expense (NRE), hurting the low volume manufacturers the most.

Recently an Interposer-based-design methodology is being investigated. It is an innovation in packaging, System on Interconnect Fabric (SoIF), that enables integration of multiple bare heterogeneous dies (or chiplets) on a interconnect pitch of 40-50 um~\cite{ssiyer20153d}. SoIF is very attractive for a low-volume manufacturer due to following advantages: 1) It speeds up the crucial design-to-market product cycle by bypassing the time-consuming verification and fabrication steps. A faster time to market gives an edge over the competitors. 2) It reduces the NRE cost through reuse of chiplets. This paves the way for a modular system design. 

The benefits of black-boxing the verification and fabrication stages have an associated security trade-off. Most of the hardware security and anti-piracy
measures intercept the design methodology at high-level synthesis, verification and fabrication stages. With these stages being black-boxed in interposer based design methodology there is a need for a different approach to meet the hardware security challenges. 

Outsourcing of fabrication stage alone by fabless SoC designers has created an opportunity for rogue elements in the supply chain to corrupt the design. SoIF designs procure chiplets that are synthesized, fabricated, tested at various design houses. This makes the supply chain more susceptible to rogue elements. Not withstanding its benefits the SoIF methodology has the twin disadvantage of increased susceptibility and disabled defenses.

We address these concerns with the following contribution:

\begin{itemize}
\item We analyze the security and privacy challenge specific to SoIF designs. To the best of our knowledge, this is the first work describing security challenges of SoIF methodology.
\item We propose the use of functionally diverse chiplets to detect any presence of Hardware Trojans.
\item We propose the use of redundancy in functionally diverse chiplets to protect IP privacy.
\item The proposals are implemented and analysis provided on a benchmark of a generic core (MIPS), a communication core (DCT), and a security core (AES). 
\end{itemize}

The rest of the paper is organized as in Sec.~\ref{sec:background} we discuss briefly the requisite background and previous work. In Sec.~\ref{sec:threat} we describe the threat model and analyze the limitations of known measures. In Sec.~\ref{sec:secpriv} we propose functionally diverse chiplets based measures that overcomes the limitations specific to SoIF methodology. Further, in Sec.~\ref{sec:result} we show the implementation of the proposed methods. Sec.~\ref{sec:conclusion} concludes the paper.

\section{Background}
\label{sec:background}
\begin{figure}[ht]
\centering
\includegraphics[width=0.45\textwidth]{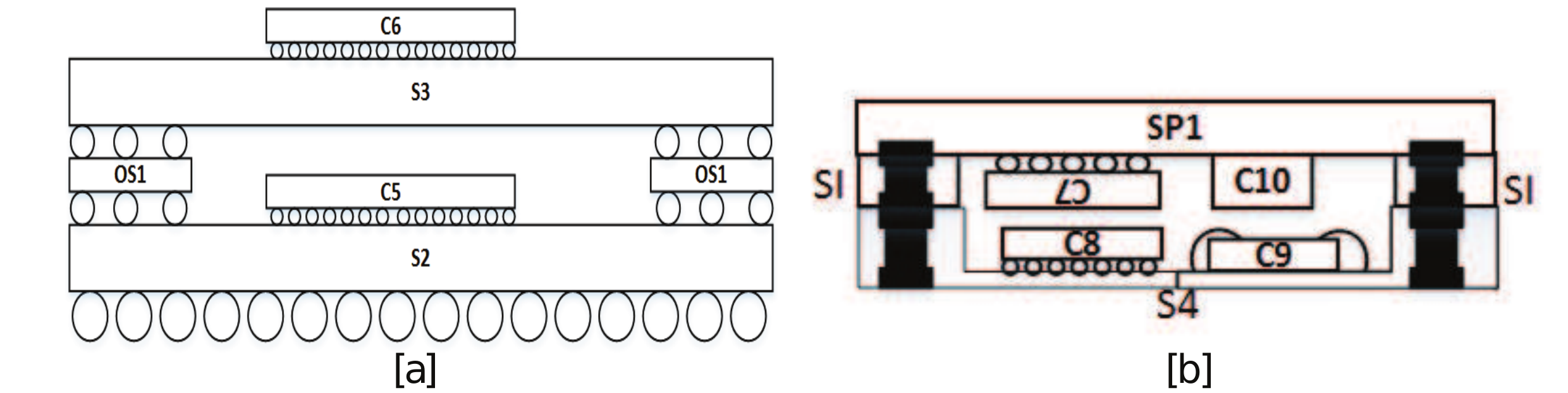}
\caption{a) A package-on-package with dies mounted on organic substrate b) Interposer with a cavity, standoff interposer and a silicon cap.}
\label{fig:interposer}
\end{figure}

Traditionally, monolithic dies of multiple electronic components are attached to a substrate that is in turn mounted on a printed circuit board (PCB). This way the finer geometries of the die are translated to match the larger spacing on a PCB. Such packages have yield concerns due to mechanical stress and lower packaging densities. Package-on-package addresses these concerns partially by stacking of dies. For example, in Fig.~\ref{fig:interposer}(a) dies \textit{C5} and \textit{C6} are connected to individual substrates \textit{S2} and \textit{S3} respectively, which are then stacked. Organic substrate \textit{OS1} provides clearance while connecting \textit{S2} and \textit{S3} electrically. This approach has a limitation on the number of stacking layers.

A recent packaging innovation is of using a silicon interposer to interconnect chiplets at much tighter pitches than PCBs~\cite{ssiyer20153d}. Fig.~\ref{fig:interposer}(b) shows an interposer with a cavity to house an electronic component and a standoff interposer to create clearances. Interposers have demonstrated heterogeneous chiplet integration and high bandwidth interconnection between chiplets~\cite{siva2017latency}.

\begin{figure}[ht]
\centering
\includegraphics[width=0.45\textwidth]{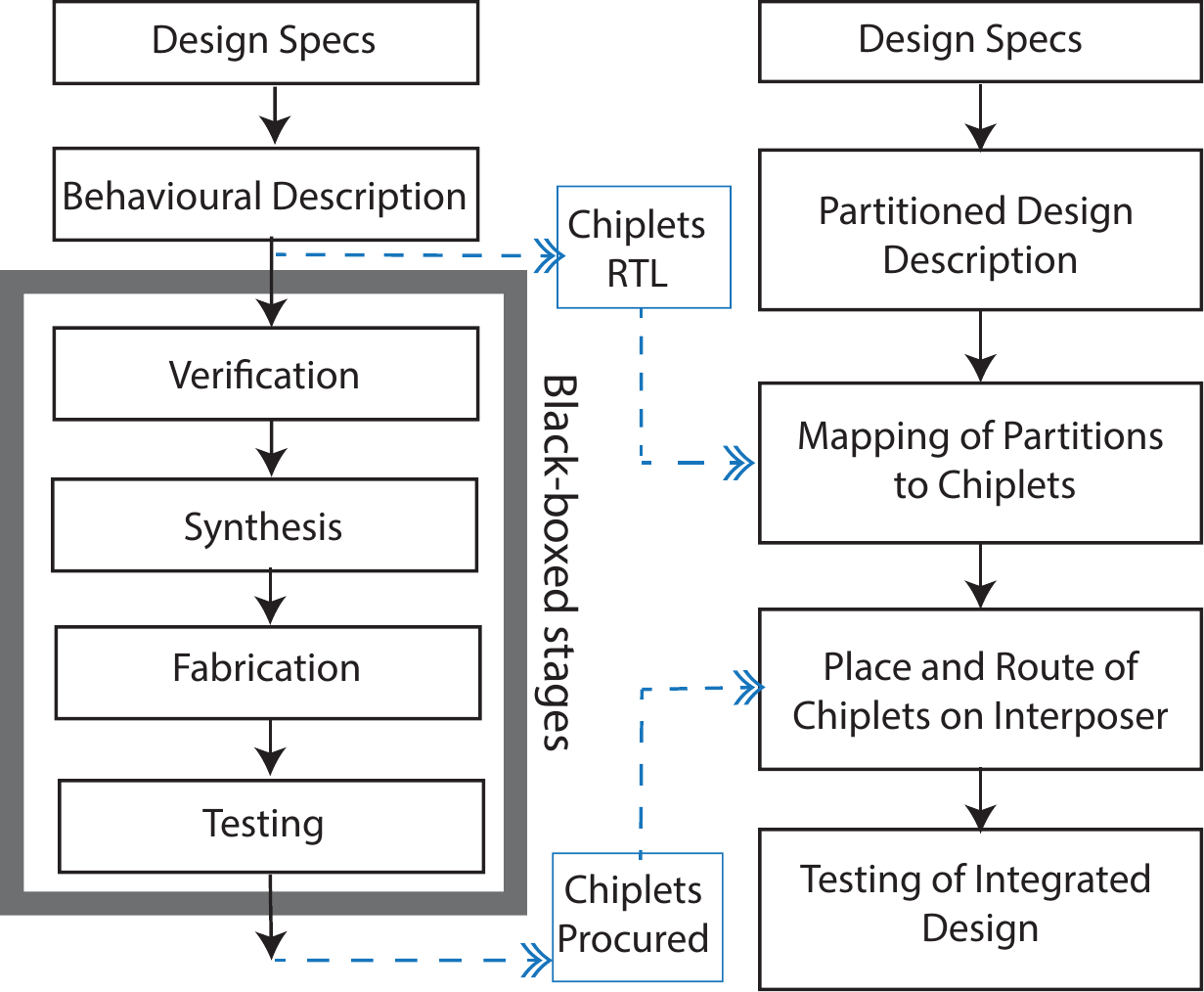}
\caption{SoC design flow using interposers to integrate chiplets. Steps on the left side are black-boxed from the integrator. Steps on the right integrator controls them.}
\label{fig:socflow_new}
\end{figure}

A SoIF integration requires partitioning of design into chiplets. The integrator procures these chiplets from different vendors from an available library of chiplets. The integrator could also procure RTL description to verify if the integrated design meets the system specifications. The design then goes through placing of chiplets and route of the interconnects. Finally, the product is tested for any defects (Fig.~\ref{fig:socflow_new}). 

Modular design allows for reuse of chiplets across different products. This brings economic viability for the chiplet vendors and chiplet integrator~\cite{daarpa}. There have been works reported in the literature to integrate IP based design to HLS \cite{sinha2014ip}, so the IP blocks can be reused at software or HDL level, saving time and effort. There also have been work to make the IP based integration Trojan aware~\cite{anir2015ip,jv2013high,jv2016high}. Such flows could be leveraged to map a required function to the chiplets in a given catalog.

\section{Threat Model} 
\label{sec:threat}
If the chiplets are procured from 3$^{rd}$ party vendors then all the upstream design stages are susceptible to mischief by rogue elements. It is possible that the chiplet vendor has in-turn procured 3$^{rd}$ Party IP (3PIP) for synthesis and/or outsourced its fabrication. Therefore one could envision several threat models
\begin{enumerate}
\item A rogue in the 3$^{rd}$ party chiplet vendor adds a Trojan into the chiplet that escapes detection by routine testing and validation. The 3$^{rd}$ party chiplet vendor could deliver a Trojan-free RTL of the IP without the Trojan. Such 3$^{rd}$ party chiplet makes it easier to sneak malicious Trojans into the design.
\item A rogue in a foundry adds Trojans into chiplets during fabrication. It is likely that this rogue may infect chiplets from different 3$^{rd}$ party vendors.
\item The 3$^{rd}$ party chiplet vendors may have sourced one or more 3PIPs. A rogue in the 3PIP vendor can add Trojans into 3PIP. The same rogue can affect multiple 3$^{rd}$ party chiplet vendors who source it from the same 3PIP vendor.
\item An attacker and designer both have equal access to Commercial-Of-The-Shelf chiplets which makes identification of chiplets by the image matching easier. The wider bump pitches of 40-50 um compared to the M1 metal width of order 144nm in an SoC, make the reverse engineering easier.
\end{enumerate}

\subsection{Previous Work}

The previous work on Trojan detection in SoC can be clubbed into three heads a) Duplication based b) RTL-verification based c) side-channel based. 

Duplication plus vendor diversity can detect a variety of Trojans in RTL 3PIPs \cite{jv2013high} \cite{jv2016high}, \cite{liu2014ip}. Each IP is duplicated and the duplicate IPs are sourced from different vendors. Its highly unlikely that two different vendors would implement the same Trojan. Therefore, it is very likely that Trojan-infected IP from different vendors produce a different result. Hence, the Trojan can be detected by comparing the results of two copies from different vendors.

The taxonomy of Trojans based on its trigger is explained as \cite{adam2011silence}: 1) Single-shot cheat code is a single data value that can trigger a Trojan. 2) Sequence cheat code is a trigger spread over multiple cycles/inputs and 3) Ticking time-bomb Trojan has a time-based trigger. RTL validation epochs are used to verify presence any Trojans \cite{adam2011silence}. Hardware modules are operated only over the number of cycles that they have been validated over all possible input combinations. Many other works also use RTL verification methods to detect Trojans \cite{adam2013fanci,jie2013veri,jie2014de} .

Side channel based detection techniques use side channel parameters like power signatures to detect extra hardware~\cite{dakshi2007fingerprint}. Hardcoding of functional assertions can also be used to detect an abnormal behavior~\cite{nuvo2010invar}. These rely on a golden model to detect an anomaly.

A survey of state-of-the-art IP piracy defense mechanism~\cite{jv2014trust} explains that the methods for regaining trust in SoC designs are: Logic encryption\cite{jarrod2008epic,jv2015logenc}, split manufacturing~\cite{jv2013split,kan2015split} and IC camouflaging~\cite{jv2014iccamou}.  Logic encryption implements hardware locking by inserting extra gates. IP can be protected by splitting the layout and manufacturing different metal layers in different foundries~\cite{jv2013split}. IC camouflaging modifies the layout of certain gates to deceive reverse engineers \cite{jv2012obfus}.

\subsection{Motivation}
Table \ref{tab:diverse} summarizes four possible ways in which diversity of vendors could be compromised in SoIF integration. This is a limitation to the current duplication plus diversity approach. Since chiplets are pre-manufactured designs, the RTL validation based approaches are ruled out for Trojan detection. 

\begin{table}[ht]
\centering
\caption{Possible diversity options}
\begin{tabular}{ |l|l|l|l| }
 \hline
 3PIP vendor & Chiplet vendor & Foundry & Diversification status\\ 
 \hline
 Different & Different  & Different & safe\\

 \hline
 Same & x	 & x & vulnerable\\ 
 \hline
x &	Same  & x & vulnerable\\
 \hline
x & x & same & vulnerable\\
 \hline
 \end{tabular}
\label{tab:diverse}

\end{table}

On the anti-piracy front, the challenges posed by chiplets include the lack of control on the placement of gates~\cite{xie2017ic}, which  restricts the ability to conceal sensitive information~\cite{jv2013split}.  Further, the obfuscation techniques~\cite{jv2012obfus} cannot apply across multiple chiplets.

\section{Security and Privacy Aware SoIF Design}
\label{sec:secpriv}
We target Trojans that are triggered by a rare event or a sequence of inputs; therefore it is not possible to detect them using routine functional tests of the chiplets. We assume use of active interposers with sparse logic and repeaters, to simplify the testing of interposers. We also assume that an interposer is a secure platform for chiplet integration. It has sparse or no logic so a thorough testing of the interposer is possible.

\subsection{Fully Diversified Duplication}
In this best-case scenario, we source functionally equivalent chiplets from diverse vendors and compare their results. This is Case A of Table~\ref{tab:diverse}. This approach diversifies all the design houses involved in the upstream of the supply chain. For malice to go undetected, rogues in the 3PIP design houses, chiplet  design houses, and the foundries may need to collude. If one of the chiplets is infected by a Trojan it can be detected by this comparison (potentially) implemented on the active interposer. Chiplets from different vendors are highly unlikely to have similar Trojans assuming no collusion.

\textit{Limitations:} This approach may be limited by the number of chiplet vendors available. This approach also makes it easy for rogues to infiltrate the supply chain either via the 3PIP vendors or via the chiplet 3$rd$ party vendors or via the foundries used to fabricate the chiplets.

\subsection{Duplication with reordered inputs}
One can apply inputs in different order to duplicate chiplets to neutralize the sequence cheat code \cite{adam2011silence}. It makes sure the same Trojan is not triggered in both the chiplets. To prevent a time-bomb trigger the chiplets are regularly refreshed.
 
\begin{figure}[ht]
\centering
\includegraphics[width=0.45\textwidth]{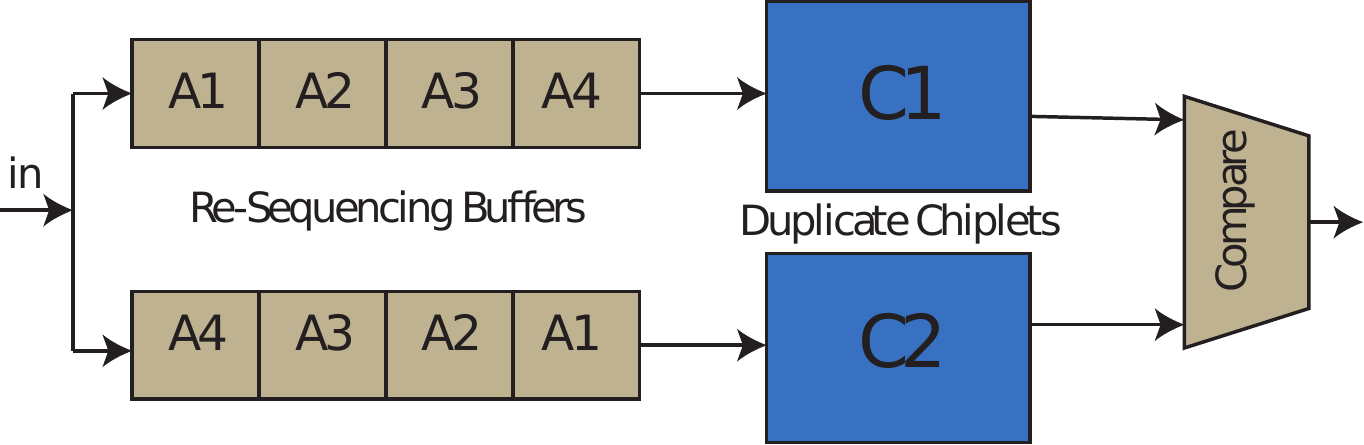}
\caption{Data flow for into buffer that reorders the inputs feeding them to the duplicated chiplets and a comparator that compares the in-order outputs}
\label{fig:chipletflow2}
\end{figure}

A reordering buffer feeds inputs to the duplicate chiplets. Their outputs are put back in order by the output buffer and the results are compared. Either the execution continues upon a match  of all outputs or an error is detected following which refresh and re-computation is done. The chiplets are periodically refreshed one after another to avoid time-based triggering. 

Since both the chiplets are fed the same inputs in different order, in the worst case the Trojan may be triggered in one of them by the cheat code leading to a detection. The ticking time-bomb Trojan is prevented by periodically refreshing the chiplets. The reorder buffer and comparator introduce extra latency and extra hardware cost of implementation for reordering inputs and outputs respectively.

\textit{Limitations:} In Fig.~\ref{fig:chipletflow2} at any given cycle both the chiplets are fed different inputs. So in the case of a single-shot cheat code trigger, Trojan would be triggered at different cycles. If the Trojan produces different outputs for the other inputs in the pipeline then Trojan can be detected. For example, if A2 is a single-shot cheat code, Trojan is triggered in cycle 2 for C1 and cycle 3 for C2. It can be detected if A3 or A4 produce different outputs in presence of a Trojan, as their output would be infected in C2 and C1 respectively. If A3 and A4 mask the Trojan then the detection fails.

\subsection{Functional Diversity}
One way to implement a function $F$ is to source a corresponding chiplet $C1$. Another way is to source a Chiplet $C2$ implementing function $G$, such that $F$ is a subset of $G$. In such case $C1$ and $C2$ are called functionally diverse chiplets. Along with diverse functions, they are likely to have diverse structure and i/o lines. This diversity makes insertion of the same Trojan in both the chiplets very difficult for the attacker. Therefore, we propose duplication by functionally diverse chiplets to detect Trojan.

\begin{figure}[ht]
\centering
\includegraphics[width=0.45\textwidth]{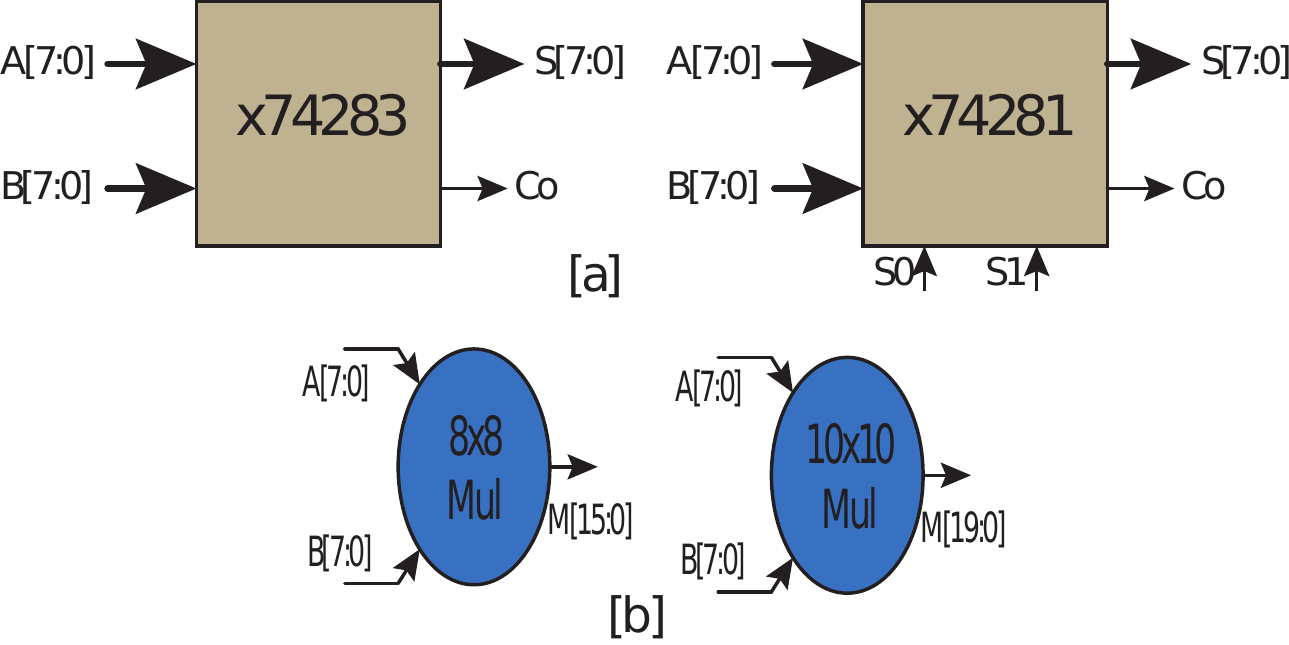}

\caption{Examples of F $\subset$ G. (a) x74281 $\subset$ x74283. (b) an 8-bit multiplier $\subset$ 10-bit multiplier.}

\label{fig:image2}
\end{figure}

The objective of an attacker is to be stealthy by making Trojan trigger a rare occurrence. This can be achieved by assigning all input lines a trigger bit and avoiding logical don't care. If attacker implements the same Trojan on functional diverse chiplets then the Trigger will have logical don't care to accommodate the diverse number of input lines. Thereby increasing the chance of Trojan detection through routine testing. This could potentially deter an attack that is susceptible to exposure.

In order to analytically capture the increase in chances of detection, we define a metric called \textit{Detection Factor (DF)} as a logarithmic function of a product of \textit{Mapping Factor (MF)} and \textit{Stealth Factor (SF)}. \textit{Mapping Factor} is the total number of possible ways of extracting function $F$ from function $G$. This depends on the number of ways data and control lines of $C1$ can be mapped to that of $C2$. \textit{Stealth Factor} is ratio how rare a trigger targeted for $F$ and $G$ can be in their input space. This depends on the total input space of $G$ and $F$ (Eq.~\ref{eq:sf}).

\begin{equation}
    DF = ln(MF*SF)
    \label{eq:df}
\end{equation}

If $C1$ has to $m$ input lines and $C2$ has $n$ input lines then \textit{SF} is given by:

\begin{equation}
    SF = \frac{\frac{1}{2^m}}{\frac{1}{2^n}} = 2^{n-m}
    \label{eq:sf}
\end{equation}

For example, let $F$ be the function of an 8-bit ALU, implemented by an ISCAS85 C880 circuit. The same can be also be implemented by C2670 (12-bit ALU and  controller). In the case of C880 and C2670, there are 4 ways of mapping the 8 input data lines onto 12 input data lines. Therefore, \textit{MF} is 4. For an attacker to implement the same Trojan in C880 and C2670 only 60 out 233 input pins in C2670 have to be used for implementing the trigger, keeping rest of 153 inputs as "don't care". Therefore, rarity of trigger in C2670 has been reduced from 1 in 2$^{233}$ to 1 in 2$^{60}$, \textit{SF} is 2$^{233}$/ 2$^{60}$. The \textit{DF}, therefore, is 175.

By diversifying the design we pay a cost of dead hardware of $C2$ that will not be used in implementing functionality $F$. So an optimal way would be to choose $C2$ such that its functionality $G$ is as diverse from $F$ while minimizing the extra hardware.  

\subsection{Diversity by distribution}
\begin{figure}[ht]
\centering
\includegraphics[width=0.45\textwidth]{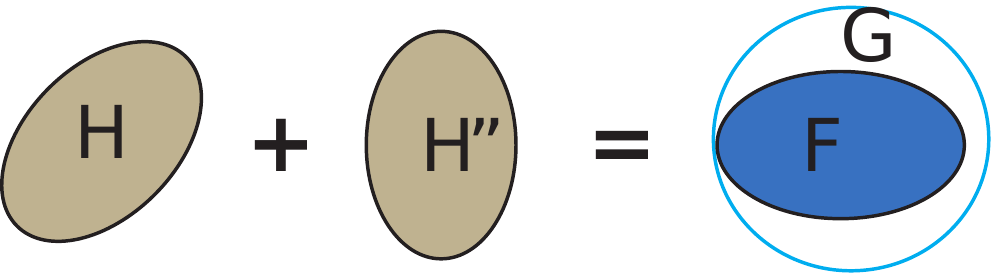}
\caption{Choosing H as subset of H".}
\label{fig:merge}
\end{figure}

Another way of implementing $F$ (corresponding to chiplet $C1$) is to distribute it over a set of smaller chiplets with sub-function of $F$ (Fig.~\ref{fig:merge}). The set of such smaller chiplets is called functionally distributed chiplets ($C3$). Functional distribution also makes it difficult to insert the same Trojan in both the chiplet $C1$ and chiplet set $C3$. We propose duplication by such functionally distributed chiplets to detect Trojan.

For an attacker to implements the same Trojan on chiplet $C1$ and $C3$ the Trojan also has to be distributed. In order to have same Trojan for function $F$ and its sub-function attacker would be forced to have logical don't cares in the trigger. This increases the chances of getting detected, which is represented by \textit{Detection Factor (DF)} (Eg.~\ref{eq:df}). 

For example, 8-bit ALU (C880) can also be implemented by using two copies of 74181 (74X series). The 8-bit input function is distributed to two 74181 chiplets, which share the control signals. The carry out of one 74181 chiplet (with lower input bits) is connected to the carry in of other 74181 chiplet (with higher input bits). \textit{MF} is only 1. In case if the same Trojan is inserted in C880 and 74181 rarity of trigger in C880 decreases from  1 in 2$^{60}$ to 1 in 2$^{14}$. \textit{SF} is 2$^{46}$ and \textit{DF} is 46.

As the logic is distributed it entails a cost of performance and number of interconnects on the interposer increases. A finer distribution into smaller sub-function would increase the cost but will be good for security.

\subsection{Ensuring Privacy}

\begin{table*}[t]
\centering
\caption{Possible ways of mapping 8-bit multiplier on 10-bit multiplier}
\begin{tabular}{ |l|l|l|l|l| }
 \hline
 Input mapping & Output mapping & key inputs & redundant input bits & redundant output bits\\ \hline
 A(7:0)$ \rightarrow $ A'(7:0) & M(15:0)$ \rightarrow $ M'(15:0) & - & A'(9:8) & M'(19:16)\\ 
 \hline
  A(7:0) $ \rightarrow $ A'(8:1) & M(15:0) $ \rightarrow $ M'(17:2) & A'(0) & A'(9) & M'(19:18,0:1)\\
 \hline
 A(7:0) $ \rightarrow $ A'(9:2) & M(15:0) $ \rightarrow  $ M'(19:4) & A'(0:1) & - & M'(0:3)\\
\hline
 \end{tabular}
\label{tab:datakeypos}

\end{table*}

In logic encryption~\cite{jarrod2008epic,jv2015logenc} extra hardware is added to deceive attacker about the functionality of the chip. The extra key gates increase the i/o space of the logic block, i.e, the encrypted circuit is a subset of the original circuit. Using the same concept we propose the use of functionally diverse chiplets to deceive an attacker about functionality of SoIF design. If function $F$ (corresponding to $C1$) can be implemented by sourcing a chiplet $C2$ that implements function $G$ if $F \subset G$. This requires the i/o space representing $G \cap (\neg F)$ to be configured for $F$. Such configuration can be treated as a ``key" similar to that of logic encryption. Two examples are discussed below:

Example for \textit{Key in the control lines}: In the Fig.~\ref{fig:image2}(a) two 74X logic blocks 74283 - adder and 74181 - ALU, be $F$ and $G$ respectively. 74181 does multiple operations on inputs depending on 5-bit control line. For a particular pattern of these 5 key bits, 74181 ($G$) does the addition operation of inputs as same as 74283 ($F$).

Example for \textit{Key in the data lines}: If the required functionality $F$ is an 8-bit multiplier and a 10-bit multiplier is sourced to implement $F$ (Fig.~\ref{fig:image2}(b)). It can be done in 3 ways as shown in the Table~\ref{tab:datakeypos}, option in the 3$^{rd}$ row gives a 4-bit key (least significant 2 bits of each operand needs to be 0). 

The more $G$ is diverse from $F$ the better is the camouflage and larger the key, this comes at the cost of extra hardware that remains unused. There is a trade off between cost and security. 

\subsubsection{Locking of the bus}
\begin{figure}[ht]
\centering
\includegraphics[width=0.45\textwidth]{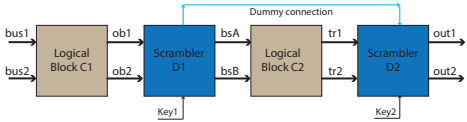}
\caption{Scrambling of interconnects between C1 and C2 using dummy blocks D1 and D2}
\label{fig:scram}
\end{figure}

Another way of protecting an IP is by locking of the bus. This can be done by use of dummy chiplets that 1) scrambles the interconnects between the chiplets with a key 2) takes dummy inputs and gives dummy outputs to confuse the attacker~\cite{jarrod2008bus}. As shown in Fig.~\ref{fig:scram} chiplets C1 and C2 integration can be logically encrypted using dummy blocks D1 and D2 that scramble the connections for key K1 and K2. Dummy connections can also be added to mislead reverse engineering.

The scrambling logic of D1 and D2 could be bit permutation implemented as Benes network~\cite{jarrod2008bus} where $n$ bits could be scrambled using $nlog_2n$ keys. This technique could also be implemented orthogonally with other logic obfuscation techniques like SARLock structure~\cite{yasin2016sar} to boost resilience.

\section{Implementation}
\label{sec:result}
We evaluate the different methods of securing chiplets discussed in the previous section. We first apply these to ISCAS circuits then later to Xilinx CORE GENERATOR IPs. And finally, apply these measures to a MIPS processor, a DCT core and AES core implementation on an interposer based FPGA.

\subsection{ISCAS benchmark}
We assume that we have a catalog of chiplets that depict ISCAS-85, 74X-Series circuits. We study examples of SoIF designs of function listed in Table~\ref{tab:method45} column 1. We implement with these functions the four Trojan detection methods discussed in Sec.~\ref{sec:secpriv} as follows: 1) Duplication by fully diversified chiplets of chiplets listed in Table~\ref{tab:method45} column 2.  2) duplication with reordered inputs of chiplets  listed in Table~\ref{tab:method45} column 2. 3) duplication by functionally diverse chiplets listed in Table~\ref{tab:method45} column 2 and 3  and 4) duplication by functionally distributed chiplets listed in Table~\ref{tab:method45} column 2 and 5.

\begin{table}[ht]
\centering
\caption{ISCAS 85 circuits possible implementations}
\label{tab:method45}
\begin{tabular}{|l|l|l|l|l|l|}
\hline
Function                                                           & Circuit & \begin{tabular}[c]{@{}l@{}}Diverse\\ Chiplets\end{tabular} & \textit{DF} & \begin{tabular}[c]{@{}l@{}}Distributed\\ Chiplets\end{tabular}  & \multicolumn{1}{c|}{\textit{DF}} \\ \hline
8-bit ALU                                                          & C3540   & C5315                                                      & 123         & 2( 74181)                                                       & 36                               \\ \hline
8-bit ALU                                                          & C880    & C2670                                                      & 175         & 2( 74181)                                                       & 46                               \\ \hline
16x16 multiplier                                                   & C6288   & -                                                          & -           & 16( S344)                                                       & 23                               \\ \hline
\begin{tabular}[c]{@{}l@{}}32-bit adder/\\ comparator\end{tabular} & C7552   & -                                                          & -           & \begin{tabular}[c]{@{}l@{}}4( 74283) +\\ 4( 74L85)\end{tabular} & 196                              \\ \hline
\end{tabular}
\end{table}

\begin{figure}[ht]
 
 \includegraphics[width=0.45\linewidth]{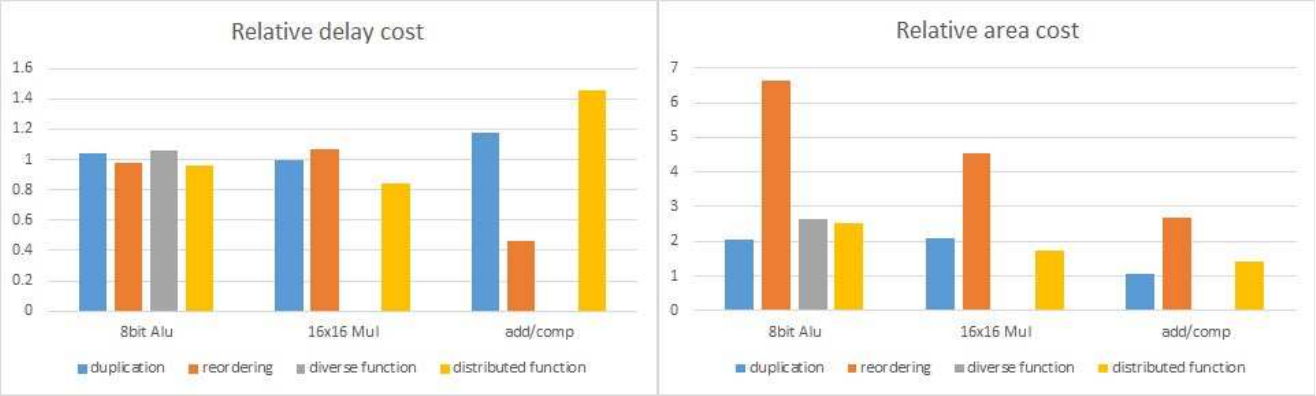}
\caption{Cost comparison of area and delay in a and  b respectively of different methods of Trojan detection discussed in Sec~\ref{sec:secpriv}.}
\label{fig:delaycost}
\end{figure}

The delay and area cost comparison of the four methods is shown in Fig.~\ref{fig:delaycost}(b) and~\ref{fig:delaycost}(a). \textit{Duplication with reordered inputs} is worst in terms of area cost as it incurs costs of inputs and outputs buffers. In \textit{duplication by functionally distributed chiplets}, the area and delay costs are better as we only break down the functionality to smaller chiplets.

\subsection{Survey of Xilinx IP catalog}
SoIF design methodology requires mapping of required function to a chiplet (Fig.~\ref{fig:socflow_new}) from a catalog. In order to emulate such mapping and to implement diversifying techniques discussed in Sec.~\ref{sec:secpriv}, we surveyed an IP catalog, Xilinx CORE GENERATOR. This is a commonly used catalog for system designs targeted to Xilinx FPGA. The possibilities of diversification of basic logic building functions IPs are listed in Table~\ref{tab:core_gen}. We further implemented these functions with two diverse IPs and Table \ref{tab:core_imp} gives its area cost and performance results.

\begin{table}[ht]
\caption{Basic Functions and Implementation Options in IP CORE GENERATOR Catalog}
\centering

\begin{tabular}{ |l|l| }
 \hline
 Functionality & Available IPs\\ \hline
 Add/Sub &  \textit{Adder-Subtracter, Accumulator, Floating-point,}\\
  & \textit{ Multiply-Adder} \\ 
 \hline
 Multiply &  \textit{Multiplier, Multiply-Accumulate, Floating-point,}\\ 
    & \textit{Multiply-Adder, Complex-multiplier}\\ \hline
 Divide &  \textit{Divider Generator, Floating-point}\\ \hline
 Sine-Cosine &  \textit{Cordic, DDS Compiler}\\ \hline
 Square-root &  \textit{Cordic, Floating-point}\\
  \hline
 \end{tabular}
\label{tab:core_gen}

\end{table}

\begin{table}[ht]
\centering
\caption{Comparison report of implementation of a given functionality on two different IPs}
\begin{tabular}{ |l|c|c|c|c| }
 \hline
 IP & LUTs & FF & Delay & Latency\\ \hline
 \multicolumn{5}{|c|}{Square Root 16-bit} \\
 \hline
 Cordic &  102 & 176 & 2.52 ns & 16 \\ 
 \hline
 Floating-point & 310 & 309 & 2.66 ns & 16 \\
 \hline
  \multicolumn{5}{|c|}{Sine Cosine 16-bit} \\
  \hline
 Cordic &  1046 & 1032 & 2.65 ns & 16 \\ 
 \hline
 DDS-compiler & 133 & 177 & 2.67 ns& 2 \\
 \hline
  \multicolumn{5}{|c|}{Divider 32-bit} \\
  \hline
 Divider-generator &  1283 & 3374 & 3.12 ns & 22 \\ 
 \hline
 Floating-point & 1283 & 1477 & 4.3 ns & 1 \\
  \hline
  \multicolumn{5}{|c|}{Multiplier 32-bit} \\
  \hline
 Multiplier &  1589 & 1869 & 3.87 ns & 3 \\ 
 \hline
 Multiply Adder & DSP48Es 4 & 130 & 1.73 ns & 3 \\
  \hline
 \end{tabular}
\label{tab:core_imp}

\end{table}

\begin{figure}[ht]
\centering
\includegraphics[width=0.45\textwidth]{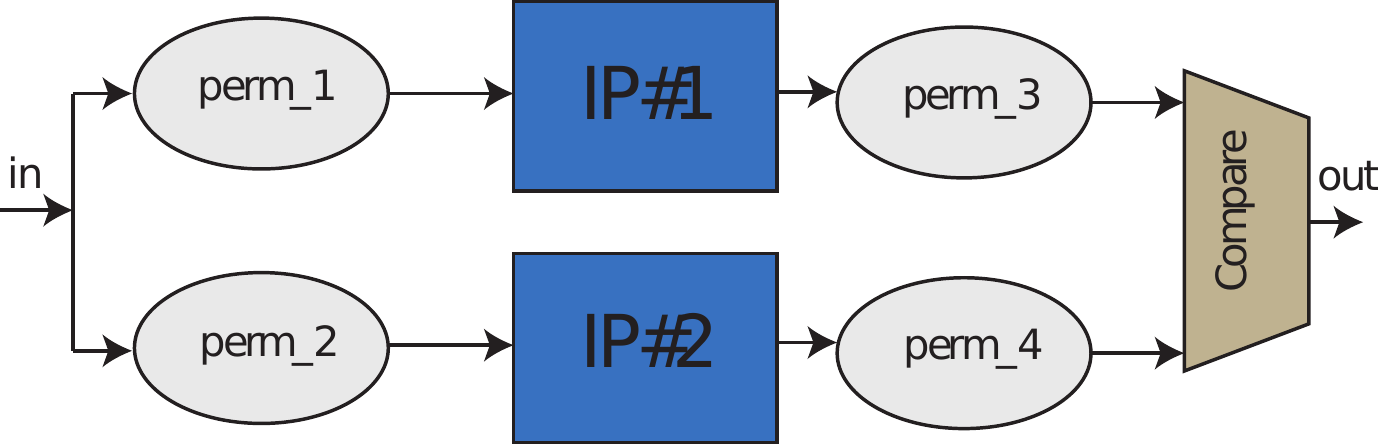}
\caption{Mapping of square root function on IP\#1 and IP\#2 with corresponding adjustments for the inputs and outputs}
\label{fig:sqrt}
\end{figure}

\textit{Security Guarantees:}  The diverse IPs apart from the difference in their implementation also have different input and output formats. Making it  very unlikely that same input vector triggers Trojan in both IPs. It is even less likely to produce the same wrong result despite being procured from the same vendor. 
Diverse IPs require permuting the inputs before applying to the IP and then permuting the outputs before comparing the results (Fig.~\ref{fig:sqrt}). For example: the square root IP implementation was done with \textit{Cordic} and \textit{Floating-point} IPs. The \textit{Cordic} IP could be used with multiple input formats we choose to implement Fix16\_14(Fix8\_6) weighting~\cite{cordic} where as \textit{Floating-point} IP operates on floating point representation~\cite{fpip}. Therefore the binary input is converted into respective formats before applying to the two IPs. The output of \textit{Cordic} IP is right shifted to get a binary representation and output of \textit{Floating-point} IP is converted back to fixed point binary  (Fig.~\ref{fig:sqrt}). This holds true irrespective of the function being linear or non-linear.

\subsection{Case Study} 

To emulate our proposal and calculate its cost we implemented a benchmark of a generic core (MIPS), a communication core (DCT) and a security core (AES). The cores were implemented on a  Virtex-7 xc7v2000tfhg176 interposer-based FPGA~\cite{virtex7}. This FPGA integrates four smaller FPGA chiplets, SLR0, SLR1, SLR2, and SLR3. We duplicated logic blocks of the core by mapping them to diverse IPs from the catalog. Such duplicated IPs were placed on different chiplets of FPGA so that the area and delay cost reflects the implementation on an interposer.

\begin{figure}[ht]
\centering
\includegraphics[width=0.45\textwidth]{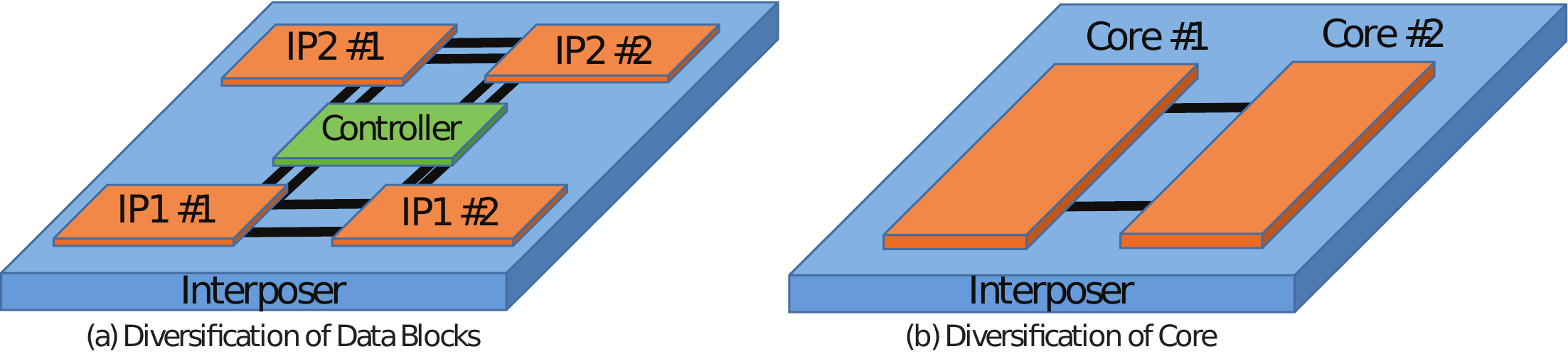}
\caption{(a) Duplication of data block IPs and a trusted controller. (b) Duplication of IP core implemented by diverse algorithms}
\label{fig:iponinterposer}
\end{figure}

One can find various ways to diversify implementation of a required function. In these case studies of MIPS core, DCT core we duplicate the blocks that handle the data, we assume that the controller is designed to support such data block implementation in a trusted way (Fig.~\ref{fig:iponinterposer}(a)). Thus we map only the data blocks to chiplets because the data blocks can be reused for different cores \cite{dylan2016cost}. In the AES core and FFT core we duplicate the whole core with different algorithmic implementation (Fig.~\ref{fig:iponinterposer}(b)). Such technique has been used in software/programming to build secure systems \cite{mcin2015}.

\subsection*{Case Study I: MIPS Core}
MIPS processor core implements the architecture described in \cite{mips}. We duplicated the divider and multiplier blocks using diverse IPs mentioned in Table~\ref{tab:core_imp}. These diverse IPs were mapped on different chiplets connected through the interposer of FPGA board. The ALU can also be duplicated as shown already in the ISCAS benchmark.

To boost anti-piracy measures bus locking was achieved through a Benes network to lock \textit{program counter} and \textit{opcode} bus by 40 key bits. These buses connect between memory and program counter and memory to multiple other blocks respectively. We further used a multiplier with wider data width (4-bit), using the extra bits as the key (8-bit). The area cost with security and privacy measures is shown in Table~\ref{mipsdet}.

\begin{table}[ht]
\centering
\caption{MIPS implementation cost analysis}
\label{mipsdet}
\begin{tabular}{|l|l|l|l|l|l|}
\hline
\textbf{}                & \textit{LUTs} & \textit{FF} & \textit{DSP} & \textit{Max Freq} & \textit{DF}\\ \hline
\textbf{duplictaed IP}   & 5150          & 5738        & 4            & 12.9ns     & 2       \\ \hline
\textbf{single IP}       & 3732          & 3420        & 0            & 12.0ns    & 0      \\ \hline
\textbf{Benes n/w}           & 3957          & 3420        & 0            & 14.1ns & 2           \\ \hline
\textbf{Benes + data redundancy} & 4163          & 3599        & 0            & 13.9ns & 3            \\ \hline
\end{tabular}
\end{table}

 \begin{figure}[ht]
\centering
\includegraphics[width=0.45\textwidth]{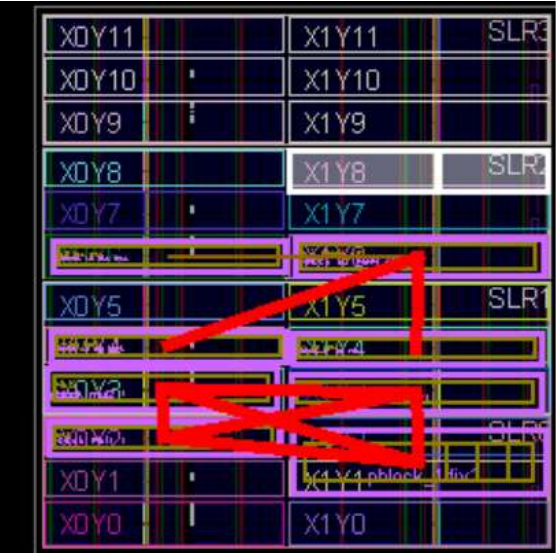}
\caption{Mapping of MIPS design on Virtex-7 xc7v2000tfhg176 FPGA board. IPs \textit{Multiplier} and \textit{Divider-Generator} are mapped to SLR0, \textit{Multiply-Adder} and \textit{Floating-point} are mapped on SLR1. Register-bank and program-counter are also mapped to SLR1. The bus-controller, Benes-key (scrambler), memory-controller are mapped to SLR2}
\label{fig:virtex}
\end{figure}

\subsection*{Case Study II: DCT Core}
Discrete Cosine Transform core is built using Multiply and Accumulate blocks (MAC). This block forms the building block along with registers and control circuit. We duplicated MACs by using diverse IPs of \textit{Multiply-Adder} and \textit{Adder-Subtracter}. The two IPs implemented were placed on different chiplets of the FPGA board. The results are shown in Table~\ref{dctaes}.

\subsection*{Case Study III; AES Core}
\begin{figure}[ht]
\centering
\includegraphics[width=0.45\textwidth]{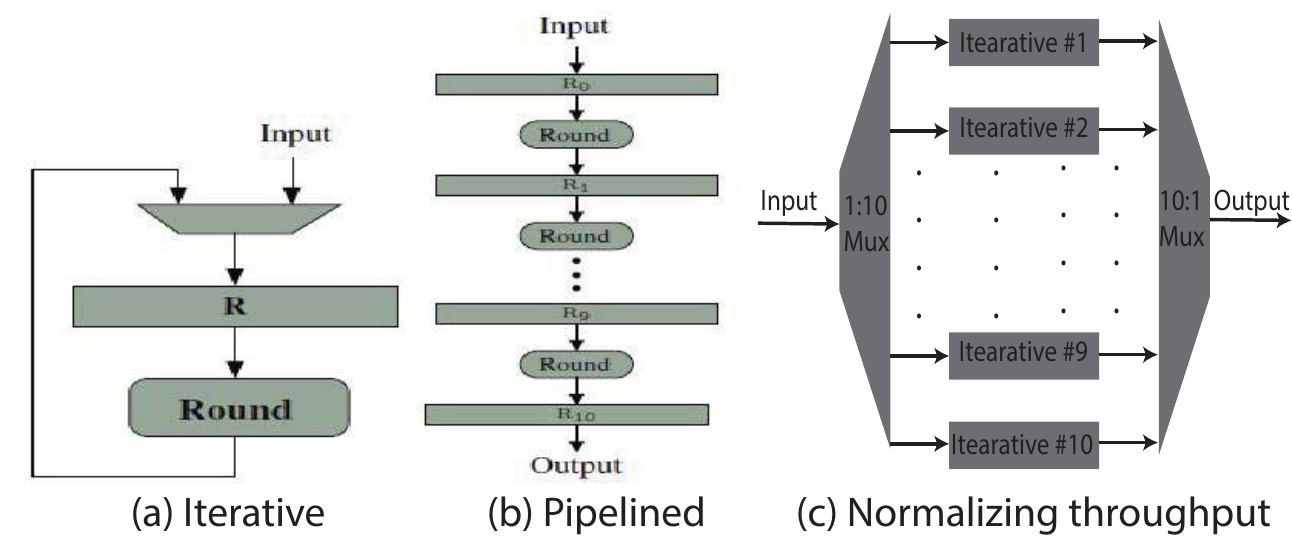}
\caption{(a) An iterative implementation of AES. (b) Pipelined implementation of AES. (c) 10 iterative AES cores connected to work on consecutive inputs to normalize throughput with pipelined implementation.}
\label{fig:aes}
\end{figure}

AES can be implemented in a pipelined fashion (Fig.~\ref{fig:aes}(b)) or in an iterative implementation (Fig.~\ref{fig:aes}(a)). In a pipelined design, multiple registers (R0 to R10 in Fig.) are used to store intermediate results. Therefore, the design provides the highest throughput. In iterative implementation each round operation is repeated for ten times  and the intermediate output is stored in a register. Iterative implementation consumes roughly one tenth of the area of the pipelined implementation, albeit with low throughput. We use 10 such iterative AES blocks to work on consecutive inputs to keep up with the throughput of pipelined architecture (Fig.~\ref{fig:aes}(c)). The duplicated IPs were mapped on different chiplets of the FPGA board. The results are shown in Table~\ref{dctaes}.

\begin{table}[ht]
\centering
\caption{DCT and AES implementation cost analysis}
\label{dctaes}
\begin{tabular}{|l|l|l|l|l|l|}
\hline
\textbf{}                & \textit{LUTs} & \textit{FF} & \textit{DSP} & \textit{Max Freq} & \textit{DF} \\ \hline
\textbf{DCT duplicated}   & 601          & 522        & 16            & 6.33ns    &       1  \\ \hline
\textbf{DCT with single IP}       & 385          & 472        & 0            & 3.63ns   &  0       \\ \hline
\textbf{AES duplicated}           & 81814          & 12782        & 0            & 8.65ns & 2           \\ \hline
\textbf{AES with single IP} & 40859          & 6636        & 0            & 7.85ns      &    0  \\ \hline

\end{tabular}
\end{table}

\subsection*{Case Study IV; FFT Core} 
Different FFT algorithms are used in literature that offer a trade-off between speed, computational resources and complexity. Different methods such as Radix-2 and Radix-4 are used to build diverse IPs. The duplicated IPs were mapped on different chiplets of the FPGA board. The results are shown in Table~\ref{dctaes}.

\section{Conclusion}
\label{sec:conclusion}
SoIF design methodology cuts down time and cost by black-boxing of crucial design stages, this makes the flow vulnerable to security and piracy breach. We analyzed the security and privacy challenge specific to the methodology. Solutions targeted for current SoCs are not readily applicable to SoIF. We proposed techniques that are more suitable to chiplet integration and shown their implementation. Our solution depends neither on the access to black-boxed design stages nor on the fully diversified vendor. Therefore we overcome the limitations posed by SoIF methodology.






\linespread{.9}
\normalsize
\bibliographystyle{IEEEtran}
\bibliography{citations}

\end{document}